
\magnification \magstep1
\raggedbottom
\openup 4\jot
\voffset6truemm
\def\cstok#1{\leavevmode\thinspace\hbox{\vrule\vtop{\vbox{\hrule\kern1pt
\hbox{\vphantom{\tt/}\thinspace{\tt#1}\thinspace}}
\kern1pt\hrule}\vrule}\thinspace}
\headline={\ifnum\pageno=1\hfill\else
\hfill {\it Twistors and spin-${3\over 2}$
potentials in quantum gravity} \hfill \fi}
\centerline {\bf TWISTORS AND
SPIN-${3\over 2}$ POTENTIALS}
\centerline {\bf IN QUANTUM GRAVITY}
\vskip 0.3cm
\centerline {\bf Giampiero Esposito${ }^{1,2}$ and
Giuseppe Pollifrone${ }^{1,3}$}
\vskip 0.3cm
\centerline {\it ${ }^{1}$Istituto Nazionale di Fisica Nucleare}
\centerline {\it Mostra d'Oltremare Padiglione 20, 80125
Napoli, Italy;}
\centerline {\it ${ }^{2}$Dipartimento di Scienze Fisiche}
\centerline {\it Mostra d'Oltremare Padiglione 19, 80125 Napoli,
Italy;}
\centerline {\it ${ }^{3}$Dipartimento di Fisica,
Universit\`a di Roma ``La Sapienza"}
\centerline {\it Piazzale Aldo Moro 2, 00185 Roma, Italy.}
\vskip 0.3cm
\noindent
{\bf Abstract.} Local boundary conditions involving field strengths
and the normal to the boundary, originally studied in
anti-de Sitter space-time, have been recently considered in
one-loop quantum cosmology. This paper derives the conditions
under which spin-lowering and spin-raising operators preserve these
local boundary conditions on a 3-sphere for fields of spin
$0,{1\over 2},1,{3\over 2}$ and $2$. Moreover, the
two-component spinor analysis of the four potentials of the totally
symmetric and independent
field strengths for spin ${3\over 2}$ is applied to
the case of a 3-sphere boundary. It is shown that such
boundary conditions can only be imposed in a flat
Euclidean background,
for which the gauge freedom in the choice
of the potentials remains.
Alternative boundary conditions for supergravity
involving the spinor-valued 1-forms for gravitinos and
the normal to the boundary are also studied.
\vskip 1cm
\noindent
{\it In: Twistor Theory, ed. S. Huggett (New York, Marcel
Dekker, 1994)}
\vskip 100cm
\leftline {\bf 1. Introduction}
\vskip 1cm
\noindent
Recent work in the literature has studied the quantization
of gauge theories and supersymmetric field theories in the
presence of boundaries, with application to one-loop
quantum cosmology [1-9]. In particular, in the work described
in [9], two possible sets of local boundary conditions were
studied. One of these, first proposed in anti-de Sitter
space-time [10-11], involves the normal to the boundary
and Dirichlet or Neumann conditions for spin $0$, the normal
and the field for massless spin-${1\over 2}$ fermions, and the
normal and totally symmetric field strengths for spins
$1,{3\over 2}$ and $2$. Although more attention has been paid
to alternative local boundary conditions motivated by
supersymmetry, as in [2-3,8-9], the analysis of the former boundary
conditions remains of mathematical and physical interest by
virtue of its links with twistor theory [9]. The aim of
this paper is to derive the mathematical properties of
the corresponding boundary-value problems in both cases,
since these are relevant for quantum cosmology and twistor theory.

For this purpose, sections 2-3 derive the conditions under which
spin-lowering and spin-raising operators preserve local boundary
conditions involving field strengths and normals. Section 4
applies the 2-spinor form of
spin-${3\over 2}$ potentials to Riemannian
4-geometries with a 3-sphere boundary. Boundary conditions on
spinor-valued 1-forms describing gravitino fields are studied
in section 5. Concluding remarks and open problems are presented
in section 6.
\vskip 10cm
\leftline {\bf 2. Spin-lowering operators in cosmology}
\vskip 1cm
\noindent
In section 5.7 of [9], a flat Euclidean background bounded by
a 3-sphere was studied. On the bounding $S^3$, the following
boundary conditions for a spin-$s$ field were required:
$$
2^{s} \; {_{e}}n^{AA'}... \; {_{e}}n^{LL'}
\; \phi_{A...L}= \epsilon \; {\widetilde \phi}^{A'...L'}
\; \; \; \; .
\eqno (2.1)
$$
With our notation, ${_{e}}n^{AA'}$ is the Euclidean normal
to $S^3$ [3,9], $\phi_{A...L}=\phi_{(A...L)}$
and ${\widetilde \phi}_{A'...L'}
={\widetilde \phi}_{(A'...L')}$ are totally symmetric
and independent (i.e. not related by any conjugation)
field strengths, which reduce to the massless
spin-${1\over 2}$ field for $s={1\over 2}$. Moreover,
the complex scalar field $\phi$ is such that its real
part obeys Dirichlet conditions on $S^3$ and its imaginary
part obeys Neumann conditions on $S^3$, or the other way
around, according to the value of the parameter
$\epsilon \equiv \pm 1$ occurring in (2.1), as described
in [9].

In flat Euclidean 4-space, we write the solutions of the
twistor equations [9,12]
$$
D_{A'}^{\; \; \;  (A} \; \omega^{B)}=0
\; \; \; \; ,
\eqno (2.2)
$$
$$
D_{A}^{\; \; (A'}\; {\widetilde \omega}^{B')}=0
\; \; \; \; ,
\eqno (2.3)
$$
as [9]
$$
\omega^{A}=(\omega^{o})^{A}-i\Bigr({_{e}}x^{AA'}\Bigr)
\pi_{A'}^{o}
\; \; \; \; ,
\eqno (2.4)
$$
$$
{\widetilde \omega}^{A'}=({\widetilde \omega}^{o})^{A'}
-i\Bigr({_{e}}x^{AA'}\Bigr){\widetilde \pi}_{A}^{o}
\; \; \; \; .
\eqno (2.5)
$$
Note that, since unprimed and primed spin-spaces are no longer
isomorphic in the case of Riemannian 4-metrics, Eq. (2.3) is not
obtained by complex conjugation of Eq. (2.2). Hence the spinor
field ${\widetilde \omega}^{B'}$ is independent of
$\omega^{B}$. This leads to distinct solutions (2.4)-(2.5), where the
spinor fields $\omega_{A}^{o},{\widetilde \omega}_{A'}^{o},
{\widetilde \pi}_{A}^{o},\pi_{A'}^{o}$
are covariantly constant with respect to the
flat connection $D$, whose corresponding spinor covariant
derivative is here denoted by $D_{AB'}$.
The following theorem can be now proved:
\vskip 0.3cm
\noindent
{\bf Theorem 2.1} Let $\omega^{D}$ be a solution of the twistor
equation (2.2) in flat Euclidean space with a 3-sphere boundary,
and let ${\widetilde \omega}^{D'}$ be the solution of the
independent equation (2.3) in the same 4-geometry with boundary.
Then a form exists of the spin-lowering operator which preserves
the local boundary conditions on $S^3$:
$$
4 \; {_{e}n^{AA'}} \; {_{e}n^{BB'}} \; {_{e}n^{CC'}}
\; {_{e}n^{DD'}} \; \phi_{ABCD}
= \epsilon \; {\widetilde \phi}^{A'B'C'D'}
\; \; \; \; ,
\eqno (2.6)
$$
$$
2^{3\over 2} \; {_{e}n^{AA'}} \; {_{e}n^{BB'}} \;
{_{e}n^{CC'}} \; \phi_{ABC} = \epsilon \; {\widetilde \phi}^{A'B'C'}
\; \; \; \; .
\eqno (2.7)
$$
Of course, the independent field strengths appearing in
(2.6)-(2.7) are assumed to satisfy the corresponding
massless free-field equations.
\vskip 0.3cm
\noindent
{\bf Proof.} Multiplying both sides of (2.6) by
${_{e}n_{FD'}}$ one gets
$$
-2 \; {_{e}n^{AA'}} \; {_{e}n^{BB'}} \;
{_{e}n^{CC'}} \; \phi_{ABCF}= \epsilon \;
{\widetilde \phi}^{A'B'C'D'} \; {_{e}n_{FD'}}
\; \; \; \; .
\eqno (2.8)
$$
Taking into account the total symmetry of the field strengths,
putting $F=D$ and multiplying both sides of (2.8) by
$\sqrt{2} \; \omega^{D}$ one finally gets
$$
-2^{3\over 2} \; {_{e}n^{AA'}} \; {_{e}n^{BB'}} \;
{_{e}n^{CC'}} \; \phi_{ABCD} \; \omega^{D}
=\epsilon \; \sqrt{2} \; {\widetilde \phi}^{A'B'C'D'}
\; {_{e}n_{DD'}} \; \omega^{D}
\; \; \; \; ,
\eqno (2.9)
$$
$$
2^{3\over 2} \; {_{e}n^{AA'}} \; {_{e}n^{BB'}} \;
{_{e}n^{CC'}} \; \phi_{ABCD} \; \omega^{D}
= \epsilon \; {\widetilde \phi}^{A'B'C'D'} \;
{\widetilde \omega}_{D'}
\; \; \; \; ,
\eqno (2.10)
$$
where (2.10) is obtained by inserting into (2.7) the definition
of the spin-lowering operator. The comparison of (2.9) and (2.10)
yields the preservation condition
$$
\sqrt{2} \; {_{e}n_{DA'}} \; \omega^{D}
=-{\widetilde \omega}_{A'}
\; \; \; \; .
\eqno (2.11)
$$
In the light of (2.4)-(2.5), equation (2.11) is found to imply
$$
\sqrt{2} \; {_{e}n_{DA'}} \; (\omega^{o})^{D}
-i \sqrt{2} \; {_{e}n_{DA'}} \; {_{e}x^{DD'}} \;
=-{\widetilde \omega}_{A'}^{o}
-i \; {_{e}x_{DA'}} \;
({\widetilde \pi}^{o})^{D}
\; \; \; \; .
\eqno (2.12)
$$
Requiring that (2.12) should be identically satisfied, and using
the identity ${_{e}n^{AA'}}={1\over r} \; {_{e}x^{AA'}}$ on a
3-sphere of radius $r$, one finds
$$
{\widetilde \omega}_{A'}^{o}=i \sqrt{2} \; r \;
{_{e}n_{DA'}} \; {_{e}n^{DD'}} \; \pi_{D'}^{o}
=-{ir \over \sqrt{2}} \; \pi_{A'}^{o}
\; \; \; \; ,
\eqno (2.13)
$$
$$
-\sqrt{2} \; {_{e}n_{DA'}} \; (\omega^{o})^{D}
=ir \; {_{e}n_{DA'}} \; ({\widetilde \pi}^{o})^{D}
\; \; \; \; .
\eqno (2.14)
$$
Multiplying both sides of (2.14) by ${_{e}n^{BA'}}$, and then
acting with $\epsilon_{BA}$ on both sides of the resulting
relation, one gets
$$
\omega_{A}^{o}=-{ir \over \sqrt{2}} \;
{\widetilde \pi}_{A}^{o}
\; \; \; \; .
\eqno (2.15)
$$
The equations (2.11), (2.13) and (2.15) completely solve the
problem of finding a spin-lowering operator which preserves
the boundary conditions (2.6)-(2.7) on $S^{3}$. Q.E.D.

If one requires local boundary conditions on $S^{3}$ involving
field strengths and normals also for lower spins (i.e. spin
${3\over 2}$ vs spin $1$, spin $1$ vs spin ${1\over 2}$, spin
${1\over 2}$ vs spin $0$), then by using the same technique of the
theorem just proved, one finds that the preservation condition obeyed
by the spin-lowering operator is still expressed
by (2.13) and (2.15).
\vskip 1cm
\leftline {\bf 3. Spin-raising operators in cosmology}
\vskip 1cm
\noindent
To derive the corresponding preservation condition for
spin-raising operators [12], we begin by studying the
relation between spin-${1\over 2}$ and spin-$1$ fields.
In this case, the independent spin-$1$ field strengths
take the form [9,11-12]
$$
\psi_{AB}=i \; {\widetilde \omega}^{L'}
\Bigr(D_{BL'} \; \chi_{A}\Bigr)
-2\chi_{(A} \; {\widetilde \pi}_{B)}^{o}
\; \; \; \; ,
\eqno (3.1)
$$
$$
{\widetilde \psi}_{A'B'}=-i \; \omega^{L}
\Bigr(D_{LB'} \; {\widetilde \chi}_{A'}\Bigr)
-2{\widetilde \chi}_{(A'} \; \pi_{B')}^{o}
\; \; \; \; ,
\eqno (3.2)
$$
where the independent spinor fields $\Bigr(\chi_{A},
{\widetilde \chi}_{A'}\Bigr)$ represent a massless
spin-${1\over 2}$ field obeying the Weyl equations
on flat Euclidean 4-space and subject to the boundary
conditions
$$
\sqrt{2} \; {_{e}}n^{AA'} \; \chi_{A}=
\epsilon \; {\widetilde \chi}^{A'}
\eqno (3.3)
$$
on a 3-sphere of radius $r$. Thus, by requiring that (3.1)
and (3.2) should obey (2.1) on $S^3$ with $s=1$, and bearing
in mind (3.3), one finds
$$ \eqalignno{
2\epsilon \biggr[\sqrt{2} \; {\widetilde \pi}_{A}^{o} \;
{\widetilde \chi}^{(A'} \; {_{e}}n^{AB')}
-{\widetilde \chi}^{(A'} \; \pi^{o \; B')}\biggr]
&=i \biggr[2 \; {_{e}}n^{AA'} \; {_{e}}n^{BB'} \;
{\widetilde \omega}^{L'} \; D_{L'(B} \; \chi_{A)}\cr
&+\epsilon \; \omega^{L} \; D_{L}^{\; \; (B'} \;
{\widetilde \chi}^{A')}\biggr]
&(3.4)\cr}
$$
on the bounding $S^3$. It is now clear how to carry out the
calculation for higher spins. Denoting by $s$ the spin
obtained by spin-raising, and defining $n \equiv 2s$,
one finds
$$ \eqalignno{
n \epsilon \biggr[\sqrt{2} \; {\widetilde \pi}_{A}^{o} \;
{_{e}}n^{A(A'} \; {\widetilde \chi}^{B'...K')}
-{\widetilde \chi}^{(A'...D'} \; \pi^{o \; K')}\biggr]
&=i \biggr[2^{n\over 2} \; {_{e}}n^{AA'} ...
{_{e}}n^{KK'} \; {\widetilde \omega}^{L'} \;
D_{L'(K} \; \chi_{A...D)}\cr
&+\epsilon \; \omega^{L} \;
D_{L}^{\; \; (K'} \; {\widetilde \chi}^{A'...D')}
\biggr]
&(3.5)\cr}
$$
on the 3-sphere boundary. In the comparison spin-$0$ vs
spin-${1\over 2}$, the preservation condition is not
obviously obtained from (3.5). The desired result is here found
by applying the spin-raising operators [12] to the
independent scalar fields $\phi$ and $\widetilde \phi$
(see below)
and bearing in mind (2.4)-(2.5) and the boundary conditions
$$
\phi = \epsilon \; {\widetilde \phi}
\; \; \; \; {\rm on} \; \; \; \; S^{3}
\; \; \; \; ,
\eqno (3.6)
$$
$$
{_{e}}n^{AA'}D_{AA'}\phi=-\epsilon \; {_{e}}n^{BB'}D_{BB'}
{\widetilde \phi}
\; \; \; \; {\rm on} \; \; \; \; S^{3}
\; \; \; \; .
\eqno (3.7)
$$
This leads to the following condition on $S^3$
(cf. equation (5.7.23) of [9]):
$$ \eqalignno{
0&=i\phi \biggr[{{\widetilde \pi}_{A}^{o}\over \sqrt{2}}
-\pi_{A'}^{o} \; {_{e}}n_{A}^{\; \; A'}\biggr]
-\biggr[{{\widetilde \omega}^{K'}\over \sqrt{2}}
\Bigr(D_{AK'}\phi\Bigr)
-{\omega_{A}\over 2} \; {_{e}}n_{C}^{\; \; K'}
\Bigr(D_{\; \; K'}^{C} \phi \Bigr)\biggr]\cr
&+\epsilon \; {_{e}}n_{(A}^{\; \; \; A'} \;
\omega^{B} \; D_{B)A'} \; {\widetilde \phi}
\; \; \; \; .
&(3.8)\cr}
$$
Note that, while the preservation conditions
(2.13) and (2.15) for
spin-lowering operators are purely algebraic, the
preservation conditions (3.5) and (3.8) for spin-raising
operators are more complicated, since they also involve
the value at the boundary of four-dimensional covariant derivatives
of spinor fields or scalar fields.
Two independent scalar fields have been
introduced, since the spinor fields obtained by applying
the spin-raising operators to $\phi$ and
${\widetilde \phi}$ respectively are independent as well
in our case.
\vskip 1cm
\leftline {\bf 4. Spin-${3\over 2}$ potentials in cosmology}
\vskip 1cm
\noindent
In this section we focus on the totally
symmetric field strengths $\phi_{ABC}$ and
${\widetilde \phi}_{A'B'C'}$ for spin-${3\over 2}$ fields,
and we express them in terms of their potentials, rather
than using spin-raising (or spin-lowering) operators. The
corresponding theory in Minkowski space-time (and curved
space-time) is described in [13-16], and adapted here to
the case of flat Euclidean 4-space with flat connection $D$.
It turns out that ${\widetilde \phi}_{A'B'C'}$ can then be
obtained from two potentials defined as follows. The first
potential satisfies the properties [13-16]
$$
\gamma_{A'B'}^{C}=\gamma_{(A'B')}^{C}
\; \; \; \; ,
\eqno (4.1)
$$
$$
D^{AA'} \; \gamma_{A'B'}^{C}=0
\; \; \; \; ,
\eqno (4.2)
$$
$$
{\widetilde \phi}_{A'B'C'}=D_{CC'} \; \gamma_{A'B'}^{C}
\; \; \; \; ,
\eqno (4.3)
$$
with the gauge freedom of replacing it by
$$
{\widehat \gamma}_{A'B'}^{C} \equiv \gamma_{A'B'}^{C}
+D_{\; \; B'}^{C} \; {\widetilde \nu}_{A'}
\; \; \; \; ,
\eqno (4.4)
$$
where ${\widetilde \nu}_{A'}$ satisfies the positive-helicity Weyl
equation
$$
D^{AA'} \; {\widetilde \nu}_{A'}=0
\; \; \; \; .
\eqno (4.5)
$$
The second potential is defined by the conditions [13-16]
$$
\rho_{A'}^{BC}=\rho_{A'}^{(BC)}
\; \; \; \; ,
\eqno (4.6)
$$
$$
D^{AA'} \; \rho_{A'}^{BC}=0
\; \; \; \; ,
\eqno (4.7)
$$
$$
\gamma_{A'B'}^{C}=D_{BB'} \; \rho_{A'}^{BC}
\; \; \; \; ,
\eqno (4.8)
$$
with the gauge freedom of being replaced by
$$
{\widehat \rho}_{A'}^{BC} \equiv \rho_{A'}^{BC}
+D_{\; \; A'}^{C} \; \chi^{B}
\; \; \; \; ,
\eqno (4.9)
$$
where $\chi^{B}$ satisfies the negative-helicity
Weyl equation
$$
D_{BB'} \; \chi^{B}=0
\; \; \; \; .
\eqno (4.10)
$$
Moreover, in flat Euclidean 4-space the field strength
$\phi_{ABC}$ is expressed in terms of the potential
$\Gamma_{AB}^{C'}=\Gamma_{(AB)}^{C'}$, independent
of $\gamma_{A'B'}^{C}$, as
$$
\phi_{ABC}=D_{CC'} \; \Gamma_{AB}^{C'}
\; \; \; \; ,
\eqno (4.11)
$$
with gauge freedom
$$
{\widehat \Gamma}_{AB}^{C'} \equiv \Gamma_{AB}^{C'}
+D_{\; \; B}^{C'} \; \nu_{A}
\; \; \; \; .
\eqno (4.12)
$$
Thus, if we insert (4.3) and (4.11) into the boundary
conditions (2.1) with $s={3\over 2}$, and require that
also the gauge-equivalent potentials (4.4) and (4.12)
should obey such boundary conditions on $S^3$, we
find that
$$
2^{3\over 2} \; {_{e}}n_{\; \; A'}^{A}
\; {_{e}}n_{\; \; B'}^{B}
\; {_{e}}n_{\; \; C'}^{C}
\; D_{CL'} \; D_{\; \; B}^{L'}
\; \nu_{A}=\epsilon \;
D_{LC'} \; D_{\; \; B'}^{L}
\; {\widetilde \nu}_{A'}
\eqno (4.13)
$$
on the 3-sphere. Note that, from now on (as already done in
(3.5) and (3.8)), covariant derivatives appearing in boundary
conditions are first taken on the background and then
evaluated on $S^3$.
In the case of our flat background, (4.13) is identically
satisfied since $D_{CL'} \; D_{\; \; \; B}^{L'} \; \nu_{A}$
and $D_{LC'} \; D_{\; \; B'}^{L} \; {\widetilde \nu}_{A'}$
vanish by virtue of spinor Ricci identities [17-18]. In
a curved background, however, denoting by $\nabla$ the
corresponding curved connection, and defining
$\cstok{\ }_{AB} \equiv \nabla_{M'(A}
\nabla_{\; \; \; B)}^{M'} \; , \; \cstok{\ }_{A'B'} \equiv
\nabla_{X(A'} \; \nabla_{\; \; B')}^{X}$,
since the spinor Ricci identities we need are [17]
$$
\cstok{\ }_{AB} \; \nu_{C}=\psi_{ABDC} \; \nu^{D}
-2\Lambda \; \nu_{(A} \; \epsilon_{B)C}
\; \; \; \; ,
\eqno (4.14)
$$
$$
\cstok{\ }_{A'B'} \; {\widetilde \nu}_{C'}
={\widetilde \psi}_{A'B'D'C'} \;
{\widetilde \nu}^{D'} -2 {\widetilde \Lambda}
\; {\widetilde \nu}_{(A'} \; \epsilon_{B')C'}
\; \; \; \; ,
\eqno (4.15)
$$
one finds that the corresponding boundary conditions
$$
2^{3\over 2} \; {_{e}}n_{\; \; A'}^{A}
\; {_{e}}n_{\; \; B'}^{B}
\; {_{e}}n_{\; \; C'}^{C}
\; \nabla_{CL'} \; \nabla_{\; \; \; B}^{L'}
\; \nu_{A}=\epsilon \; \nabla_{LC'}
\; \nabla_{\; \; B'}^{L}
\; {\widetilde \nu}_{A'}
\eqno (4.16)
$$
are identically satisfied if and only if one of the
following conditions holds: (i) $\nu_{A}=
{\widetilde \nu}_{A'}=0$; (ii) the Weyl spinors
$\psi_{ABCD},{\widetilde \psi}_{A'B'C'D'}$ and the
scalars $\Lambda,{\widetilde \Lambda}$ vanish everywhere.
However, since in a curved space-time
with vanishing $\Lambda,{\widetilde \Lambda}$, the potentials
with the gauge freedoms (4.4) and (4.12) only exist provided
$D$ is replaced by $\nabla$ and the trace-free part
$\Phi_{ab}$ of the Ricci tensor vanishes as well [19],
the background 4-geometry is actually flat
Euclidean 4-space. Note that we require that (4.16) should
be identically satisfied to avoid, after a gauge
transformation, obtaining more boundary conditions than
the ones originally imposed. The curvature of the background
should not, itself, be subject to a boundary condition.

The same result can be derived by using
the potential $\rho_{A'}^{BC}$ and its independent
counterpart $\Lambda_{A}^{B'C'}$. This spinor field
yields the $\Gamma_{AB}^{C'}$ potential by means of
$$
\Gamma_{AB}^{C'}=D_{BB'} \; \Lambda_{A}^{B'C'}
\; \; \; \; ,
\eqno (4.17)
$$
and has the gauge freedom
$$
{\widehat \Lambda}_{A}^{B'C'} \equiv \Lambda_{A}^{B'C'}
+D_{\; \; A}^{C'} \; {\widetilde \chi}^{B'}
\; \; \; \; ,
\eqno (4.18)
$$
where ${\widetilde \chi}^{B'}$ satisfies the positive-helicity
Weyl equation
$$
D_{BF'} \; {\widetilde \chi}^{F'}=0
\; \; \; \; .
\eqno (4.19)
$$
Thus, if also the gauge-equivalent potentials (4.9) and (4.18)
have to satisfy the boundary conditions (2.1) on $S^3$, one
finds
$$
2^{3\over 2} \; {_{e}}n_{\; \; A'}^{A}
\; {_{e}}n_{\; \; B'}^{B}
\; {_{e}}n_{\; \; C'}^{C}
\; D_{CL'} \; D_{BF'} \;
D_{\; \; A}^{L'} \;
{\widetilde \chi}^{F'}
=\epsilon \; D_{LC'} \; D_{MB'} \;
D_{\; \; A'}^{L} \; \chi^{M}
\eqno (4.20)
$$
on the 3-sphere. In our flat background, covariant derivatives
commute, hence (4.20) is identically satisfied by virtue of (4.10)
and (4.19). However, in the curved case the boundary conditions
(4.20) are replaced by
$$
2^{3\over 2} \; {_{e}}n_{\; \; A'}^{A}
\; {_{e}}n_{\; \; B'}^{B}
\; {_{e}}n_{\; \; C'}^{C}
\; \nabla_{CL'} \; \nabla_{BF'}
\; \nabla_{\; \; A}^{L'}
\; {\widetilde \chi}^{F'}
=\epsilon \; \nabla_{LC'} \;
\nabla_{MB'} \; \nabla_{\; \; A'}^{L}
\; \chi^{M}
\eqno (4.21)
$$
on $S^3$, if the {\it local} expressions of $\phi_{ABC}$ and
${\widetilde \phi}_{A'B'C'}$ in terms of potentials still
hold [13-16]. By virtue of (4.14)-(4.15), where $\nu_{C}$ is
replaced by $\chi_{C}$ and ${\widetilde \nu}_{C'}$ is
replaced by ${\widetilde \chi}_{C'}$, this means that
the Weyl spinors $\psi_{ABCD},{\widetilde \psi}_{A'B'C'D'}$
and the scalars $\Lambda,{\widetilde \Lambda}$ should
vanish, since one should find
$$
\nabla^{AA'} \; {\widehat \rho}_{A'}^{BC}=0
\; \; \; \; , \; \; \; \;
\nabla^{AA'} \; {\widehat \Lambda}_{A}^{B'C'}=0
\; \; \; \; .
\eqno (4.22)
$$
If we assume that
$\nabla_{BF'} \; {\widetilde \chi}^{F'}=0$ and
$\nabla_{MB'} \; \chi^{M}=0$, we have to show that (4.21)
differs from (4.20) by terms involving a part of the curvature
that is vanishing everywhere.
This is proved by using the basic rules
of 2-spinor calculus and spinor Ricci identities [17-18].
Thus, bearing in mind that [17]
$$
\cstok{\ }^{AB} \; {\widetilde \chi}_{B'}
=\Phi_{\; \; \; \; L'B'}^{AB} \;
{\widetilde \chi}^{L'}
\; \; \; \; ,
\eqno (4.23)
$$
$$
\cstok{\ }^{A'B'} \; \chi_{B}
={\widetilde \Phi}_{\; \; \; \; \; \; LB}^{A'B'}
\; \chi^{L}
\; \; \; \; ,
\eqno (4.24)
$$
one finds
$$ \eqalignno{
\nabla^{BB'} \; \nabla^{CA'} \; \chi_{B}&=
\nabla^{(BB'} \; \nabla^{C)A'} \; \chi_{B}
+\nabla^{[BB'} \; \nabla^{C]A'} \; \chi_{B} \cr
&=-{1\over 2} \nabla_{B}^{\; \; B'} \;
\nabla^{CA'} \; \chi^{B}
+{1\over 2} {\widetilde \Phi}^{A'B'LC} \; \chi_{L}
\; \; \; \; .
&(4.25)\cr}
$$
Thus, if ${\widetilde \Phi}^{A'B'LC}$ vanishes, also the left-hand side
of (4.25) has to vanish since this leads to the equation
$
\nabla^{BB'} \; \nabla^{CA'} \; \chi_{B}
={1\over 2}
\nabla^{BB'} \; \nabla^{CA'} \; \chi_{B}
$. Hence (4.25) is identically satisfied. Similarly, the
left-hand side of (4.21) can be made to vanish identically
provided the additional condition $\Phi^{CDF'M'}=0$ holds.
The conditions
$$
\Phi^{CDF'M'}=0
\; \; \; \; , \; \; \; \;
{\widetilde \Phi}^{A'B'CL}=0
\; \; \; \; ,
\eqno (4.26)
$$
when combined with the conditions
$$
\psi_{ABCD}={\widetilde \psi}_{A'B'C'D'}=0
\; \; \; \; , \; \; \; \;
\Lambda={\widetilde \Lambda}=0
\; \; \; \; ,
\eqno (4.27)
$$
arising from (4.22) for the local existence
of $\rho_{A'}^{BC}$ and $\Lambda_{A}^{B'C'}$ potentials,
imply that the whole Riemann curvature should vanish.
Hence, in the boundary-value problems we are interested in,
the only admissible background 4-geometry (of the Einstein
type [20]) is flat Euclidean 4-space.
\vskip 1cm
\leftline {\bf 5. Boundary conditions in supergravity}
\vskip 1cm
\noindent
The boundary conditions studied in the previous
sections are not appropriate if one studies
supergravity multiplets and supersymmetry transformations at the
boundary [9]. By contrast, it turns out
one has to impose another set of locally supersymmetric boundary
conditions, first proposed in [21]. These are in general mixed, and
involve in particular Dirichlet conditions for the transverse modes of
the vector potential of electromagnetism, a mixture of Dirichlet and
Neumann conditions for scalar fields, and local boundary conditions for
the spin-${1\over 2}$ field and the spin-${3\over 2}$ potential. Using
two-component spinor notation for supergravity [9,22], the
spin-${3\over 2}$ boundary conditions take the form
$$
\sqrt{2} \; {_{e}n_{A}^{\; \; A'}} \;
\psi_{\; \; i}^{A}=\epsilon \;
{\widetilde \psi}_{\; \; i}^{A'}
\; \; \; \; {\rm on} \; \; \; \; S^{3}
\; \; \; \; .
\eqno (5.1)
$$
With our notation, $\epsilon \equiv \pm 1$,
${_{e}n_{A}^{\; \; A'}}$ is the Euclidean normal to $S^{3}$,
and $\Bigr(\psi_{\; \; i}^{A},{\widetilde \psi}_{\; \; i}^{A'}\Bigr)$
are the {\it independent} (i.e. not related by any conjugation)
spatial components (hence $i=1,2,3$) of the spinor-valued
1-forms appearing in the action functional of Euclidean supergravity
[9,22].

It appears necessary to understand whether
the analysis in the previous section and in [23]
can be used to derive
restrictions on the classical boundary-value problem corresponding
to (5.1). For this purpose, we study a Riemannian background 4-geometry,
and we use the decompositions of the spinor-valued 1-forms in such
a background, i.e. [9]
$$
\psi_{\; \; i}^{A}= {h^{- {1 \over 4}}}
\biggr[\chi^{(AB)B'}
+\epsilon^{AB} \;
{\widetilde \phi}^{B'}\biggr]e_{BB'i}
\; \; \; \; ,
\eqno (5.2)
$$
$$
{\widetilde \psi}_{\; \; i}^{A'}
={h^{- {1 \over 4}}}
\biggr[{\widetilde \chi}^{(A'B')B}+\epsilon^{A'B'}\phi^{B}
\biggr]e_{BB'i}
\; \; \; \; ,
\eqno (5.3)
$$
where $h$ is the determinant of the 3-metric on $S^{3}$, and
$e_{BB'i}$ is the spatial component of the tetrad, written
in 2-spinor language. If we now reduce
the classical theory of simple supergravity to its physical
degrees of freedom by imposing the gauge conditions [9]
$$
e_{AA'}^{\; \; \; \; \; \; i} \; \psi_{\; \; i}^{A}=0
\; \; \; \; ,
\eqno (5.4)
$$
$$
e_{AA'}^{\; \; \; \; \; \; i} \; {\widetilde \psi}_{\; \; i}^{A'}=0
\; \; \; \; ,
\eqno (5.5)
$$
we find that the expansions of (5.2)-(5.3) on a family of 3-spheres
centred on the origin take the forms [9]
$$
\psi_{\; \; i}^{A}
={h^{-{1\over 4}} \over 2\pi}
\sum_{n=0}^{\infty}\sum_{p,q=1}^{(n+1)(n+4)}
\alpha_{n}^{pq}
\biggr[m_{np}^{(\beta)}(\tau) \; \beta^{nqABB'}
+{\widetilde r}_{np}^{(\mu)}(\tau) \;
{\overline \mu}^{nqABB'}\biggr]e_{BB'i}
\; \; \; \; ,
\eqno (5.6)
$$
$$
{\widetilde \psi}_{\; \; i}^{A'}
={h^{-{1\over 4}} \over 2\pi}
\sum_{n=0}^{\infty}\sum_{p,q=1}^{(n+1)(n+4)}
\alpha_{n}^{pq}
\biggr[{\widetilde m}_{np}^{(\beta)}(\tau)
\; {\overline \beta}^{nqA'B'B}
+ r_{np}^{(\mu)}(\tau) \;
\mu^{nqA'B'B}\biggr]e_{BB'i}
\; \; \; \; .
\eqno (5.7)
$$
With our notation,
$\alpha_{n}^{pq}$ are block-diagonal matrices with blocks
$\pmatrix {1&1 \cr 1&-1 \cr}$, and the $\beta$- and $\mu$-harmonics
on $S^{3}$ are given by [9]
$$
\beta_{\; \; \; ACC'}^{nq}=\rho_{\; \; \; (ACD)}^{nq}
\; n_{\; \; C'}^{D}
\; \; \; \; ,
\eqno (5.8)
$$
$$
\mu_{\; \; \; A'B'B}^{nq}=\sigma_{\; \; \; (A'B'C')}^{nq}
\; n_{B}^{\; \; C'}
\; \; \; \; .
\eqno (5.9)
$$
In the light of (5.6)-(5.9), one gets the following
physical-degrees-of-freedom form of the spinor-valued 1-forms
of supergravity  (cf. [9,22,24]):
$$
\psi_{\; \; i}^{A}={h^{-{1\over 4}}}
\; \phi^{(ABC)} \; {_{e}n_{C}^{\; \; B'}} \; e_{BB'i}
\; \; \; \; ,
\eqno (5.10)
$$
$$
{\widetilde \psi}_{\; \; i}^{A'}={h^{-{1\over 4}}} \;
{\widetilde \phi}^{(A'B'C')} \; {_{e}n_{\; \; C'}^{B}}
\; e_{BB'i}
\; \; \; \; ,
\eqno (5.11)
$$
where $\phi^{(ABC)}$ and ${\widetilde \phi}^{(A'B'C')}$
are totally symmetric and independent spinor fields.

Within this framework, a {\it sufficient}
condition for the validity
of the boundary conditions (5.1) on $S^{3}$ is
$$
\sqrt{2} \; {_{e}n_{A}^{\; \; A'}} \; {_{e}n_{C}^{\; \; B'}}
\; \phi^{(ABC)}= \epsilon \; {_{e}n_{\; \; C'}^{B}}
\; {\widetilde \phi}^{(A'B'C')}
\; \; \; \; .
\eqno (5.12)
$$
{}From now on, one can again try
to express {\it locally}
$\phi^{(ABC)}$ and ${\widetilde \phi}^{(A'B'C')}$ in terms of
four potentials as in section 4 and in [23],
providing they are solutions of massless free-field equations.
The alternative possibility is to consider the Rarita-Schwinger
form of the field strength, written in 2-spinor language.
The corresponding potential is no longer symmetric as in (4.1),
and is instead subject to the equations (cf. [13-16,25])
$$
\epsilon^{B'C'} \; \nabla_{A(A'} \; \gamma_{\; \; B')C'}^{A}=0
\; \; \; \; ,
\eqno (5.13)
$$
$$
\nabla^{B'(B} \; \gamma_{\; \; \; B'C'}^{A)}=0
\; \; \; \; .
\eqno (5.14)
$$
Moreover, the spinor field ${\widetilde \nu}_{A'}$
appearing in the gauge transformation (4.4) is no longer
taken to be a solution of the positive-helicity Weyl
equation (4.5). Hence the classical boundary-value
problem might have new features with respect
to the analysis of section 4 and [23].

Indeed, the investigation appearing in this
section is incomplete, and it relies in part on the
unfinished work in [26].
Moreover, it should be emphasized that our
analysis, although motivated
by quantum cosmology, is entirely classical. Hence we have not
discussed ghost modes.
The theory has been reduced to its physical degrees of freedom
to make a comparison with the results in [23], but totally
symmetric field strengths do not enable one to recover the full
physical content of simple supergravity. Hence the 4-sphere
background studied in [2] is not ruled out by our work [26].
\vskip 1cm
\leftline {\bf 6. Results and open problems}
\vskip 1cm
\noindent
Following [9] and [23], we have derived the conditions
(2.13), (2.15), (3.5), and (3.8)
under which spin-lowering and spin-raising
operators preserve the local boundary conditions studied in
[9-11]. Note that, for spin $0$, we have introduced a pair of
independent scalar fields on the real Riemannian section of
a complex space-time, following [27], rather than a single scalar
field, as done in [9]. Setting $\phi \equiv \phi_{1}+i\phi_{2},
{\widetilde \phi} \equiv \phi_{3}+i\phi_{4}$, this choice leads
to the boundary conditions
$$
\phi_{1}=\epsilon \; \phi_{3}
\; \; \; \;
\phi_{2}=\epsilon \; \phi_{4}
\; \; \; \; {\rm on}
\; \; \; \; S^{3}
\; \; \; \; ,
\eqno (6.1)
$$
$$
{_{e}}n^{AA'} \; D_{AA'} \; \phi_{1}=-\epsilon \;
{_{e}}n^{AA'} \; D_{AA'} \; \phi_{3}
\; \; \; \; {\rm on}
\; \; \; \; S^{3}
\; \; \; \; ,
\eqno (6.2)
$$
$$
{_{e}}n^{AA'} \; D_{AA'} \; \phi_{2}=-\epsilon \;
{_{e}}n^{AA'} \; D_{AA'} \; \phi_{4}
\; \; \; \; {\rm on}
\; \; \; \; S^{3}
\; \; \; \; ,
\eqno (6.3)
$$
and it deserves further study.

We have then focused on the potentials for
spin-${3\over 2}$ field strengths in
flat or curved Riemannian 4-space bounded
by a 3-sphere. Remarkably, it turns out that
local boundary conditions involving field strengths and
normals can only be imposed in a flat Euclidean background,
for which the gauge freedom in the choice of the
potentials remains. In [16] it was found that $\rho$ potentials
exist {\it locally} only in the self-dual Ricci-flat case,
whereas $\gamma$ potentials may be introduced in the
anti-self-dual case.
Our result may be interpreted as a further restriction provided
by (quantum) cosmology. What happens is that the boundary
conditions (2.1) fix at the boundary a spinor field involving
{\it both} the field strength $\phi_{ABC}$ and the field
strength ${\widetilde \phi}_{A'B'C'}$. The local existence
of potentials for the field strength $\phi_{ABC}$, jointly
with the occurrence of a boundary, forces half of the Riemann
curvature of the background to vanish. Similarly, the remaining
half of such Riemann curvature has to vanish on considering the
field strength ${\widetilde \phi}_{A'B'C'}$. Hence the
background 4-geometry can only be flat Euclidean space. This
is different from the analysis in [13-16], since in that case
one is not dealing with boundary conditions forcing us to
consider both $\phi_{ABC}$ and ${\widetilde \phi}_{A'B'C'}$.

A naturally occurring question is whether the potentials studied
in this paper can be used to perform one-loop calculations for
spin-${3\over 2}$ field strengths subject to (2.1) on $S^3$.
This problem may provide another example (cf. [9]) of the fertile
interplay between twistor theory and quantum cosmology [26], and its
solution might shed new light on one-loop quantum cosmology
and on the quantization program for gauge theories in the presence
of boundaries [1-9]. For this purpose, as shown in recent
papers by ourselves and other co-authors [28-30], it is necessary
to study Riemannian background 4-geometries bounded by two
concentric 3-spheres (cf. sections 2-5). Moreover, the consideration
of non-physical degrees of freedom of gauge fields,
set to zero in our classical analysis, is necessary
to achieve a covariant quantization scheme.
\vskip 10cm
\leftline {\bf Acknowledgments}
\vskip 1cm
\noindent
We are indebted to Stephen Huggett for suggesting
we should prepare our contribution to this volume,
and to Roger Penrose for bringing Refs. [13-16] to our attention.
Many conversations with Alexander Yu. Kamenshchik on one-loop
quantum cosmology and related problems have stimulated our
research.
\vskip 1cm
\leftline {\bf References}
\vskip 1cm
\item {[1]}
Moss I. G. and Poletti S. (1990) {\it Nucl. Phys.}
B {\bf 341}, 155.
\item {[2]}
Poletti S. (1990) {\it Phys. Lett.} {\bf 249B}, 249.
\item {[3]}
D'Eath P. D. and Esposito G. (1991) {\it Phys. Rev.}
D {\bf 43}, 3234.
\item {[4]}
D'Eath P. D. and Esposito G (1991) {\it Phys. Rev.}
D {\bf 44}, 1713.
\item {[5]}
Barvinsky A. O., Kamenshchik A. Yu., Karmazin I. P. and
Mishakov I. V. (1992) {\it Class. Quantum Grav.}
{\bf 9}, L27.
\item {[6]}
Kamenshchik A. Yu. and Mishakov I. V. (1992)
{\it Int. J. Mod. Phys.} A {\bf 7}, 3713.
\item {[7]}
Barvinsky A. O., Kamenshchik A. Yu. and Karmazin I. P. (1992)
{\it Ann. Phys., N.Y.} {\bf 219}, 201.
\item {[8]}
Kamenshchik A. Yu. and Mishakov I. V. (1993) {\it Phys. Rev.}
D {\bf 47}, 1380.
\item {[9]}
Esposito G. (1994) {\it Quantum Gravity, Quantum Cosmology
and Lorentzian Geometries} Lecture Notes in Physics,
New Series m: Monographs vol m12
second corrected and enlarged edition
(Berlin: Springer).
\item {[10]}
Breitenlohner P. and Freedman D. Z. (1982) {\it Ann. Phys., N.Y.}
{\bf 144}, 249.
\item {[11]}
Hawking S. W. (1983) {\it Phys. Lett.} {\bf 126B}, 175.
\item {[12]}
Penrose R. and Rindler W. (1986) {\it Spinors and Space-Time,
Vol. 2: Spinor and Twistor Methods in Space-Time Geometry}
(Cambridge: Cambridge University Press).
\item {[13]}
Penrose R. (1990) {\it Twistor Newsletter} {\bf n 31}, 6.
\item {[14]}
Penrose R. (1991) {\it Twistor Newsletter} {\bf n 32}, 1.
\item {[15]}
Penrose R. (1991) {\it Twistor Newsletter} {\bf n 33}, 1.
\item {[16]}
Penrose R. (1991) Twistors as Spin-${3\over 2}$ Charges
{\it Gravitation and Modern Cosmology} eds A. Zichichi,
V. de Sabbata and N. S\'anchez (New York: Plenum Press).
\item {[17]}
Ward R. S. and Wells R. O. (1990) {\it Twistor Geometry and
Field Theory} (Cambridge: Cambridge University Press).
\item {[18]}
Esposito G. (1993) {\it Nuovo Cimento} B {\bf 108}, 123.
\item {[19]}
Buchdahl H. A. (1958) {\it Nuovo Cim.} {\bf 10}, 96.
\item {[20]}
Besse A. L. (1987) {\it Einstein Manifolds} (Berlin: Springer).
\item {[21]}
Luckock H. C. and Moss I. G. (1989) {\it Class. Quantum Grav.}
{\bf 6}, 1993.
\item {[22]}
D'Eath P. D. (1984) {\it Phys. Rev.} D {\bf 29}, 2199.
\item {[23]}
Esposito G. and Pollifrone G. (1994) {\it Class. Quantum Grav.}
{\bf 11}, 897.
\item {[24]}
Sen A. (1981) {\it J. Math. Phys.} {\bf 22}, 1781.
\item {[25]}
Esposito G. (1994) {\it Complex General Relativity}
(book in preparation).
\item {[26]}
Esposito G., Kamenshchik A. Yu., Mishakov I. V. and
Pollifrone G. (1994) {\it Supersymmetric Boundary Conditions
in Quantum Cosmology}, work in progress.
\item {[27]}
Hawking S. W. (1979) The path integral approach to quantum gravity
{\it General Relativity, an Einstein Centenary Survey}
eds S. W. Hawking and W. Israel (Cambridge: Cambridge University
Press).
\item {[28]}
Esposito G. (1994) {\it Class. Quantum Grav.} {\bf 11}, 905.
\item {[29]}
Esposito G., Kamenshchik A. Yu., Mishakov I. V. and
Pollifrone G. (1994) {\it Euclidean Maxwell Theory in the
Presence of Boundaries, Part II}, DSF preprint 94/4.
\item {[30]}
Esposito G., Kamenshchik A. Yu., Mishakov I. V. and
Pollifrone G. (1994) {\it Gravitons in One-Loop Quantum
Cosmology: Correspondence Between Covariant and Non-Covariant
Formalisms}, DSF preprint 94/14.

\bye